\documentclass {aastex62}

\received{September 14, 2018}
\revised{ }
\accepted{ }

\shorttitle{Binaries and the cluster mass}
\shortauthors{Borodina et al.}

\begin{document}

\title{Unresolved binaries and Galactic clusters' mass estimates}

\correspondingauthor{Anton F. Seleznev}
\email{anton.seleznev@urfu.ru}

\author{Olga I. Borodina}
\affil{Ural Federal University \\
620002, 19 Mira street, \\
Ekaterinburg, Russia}

\author{Anton F. Seleznev}
\affil{Ural Federal University \\
620002, 19 Mira street, \\
Ekaterinburg, Russia}

\author{Giovanni Carraro}
\affil{Dipartimento di Fisica e Astronomia, Universita'  di Padova\\
Vicolo Osservatorio 3\\
I35122, Padova, Italy}

\author{Vladimir M. Danilov}
\affil{Ural Federal University \\
620002, 19 Mira street, \\
Ekaterinburg, Russia}

\begin{abstract}
Binary stars are present in all stellar systems, yet their role is far from being fully understood.
We investigate the effect of unresolved binaries in the derivation of open clusters' mass by star counts.
We start from the luminosity functions
of five open clusters: IC 2714, NGC 1912, NGC 2099, NGC 6834 and NGC 7142. Luminosity functions are
obtained via star counts extracted from  the 2MASS database. The fraction of binaries
is considered to be independent on stellar magnitude. We take into account
different assumptions for the binary mass ratio distribution and assign binary masses
using the so-called {\it luminosity-limited pairing}  method and Monte-Carlo simulations.  We show that
cluster masses increase when binary stars are appropriately taken into account.
\end{abstract}


\section{Introduction}
\citet{HH}  provided one of the first indications that star clusters harbor a large number of unresolved binary stars.
\citet{Maeder} showed what position  binary
stars have in the color-magnitude diagram (CMD) as a function of their mass ratio $q=M_2/M_1$ (where $M_2$ is the mass of the secondary
while $M_1$ is the mass of the primary component). \citet{HT} demonstrated that the secondary sequences routinely seen above the main sequence (MS)
in clusters' CMDs are actually made of binaries with wide mass ratio ranges (and not merely by equal mass binaries).
A summary of the results on the binary stars content of star clusters is presented by \citet{DK}.

The binary fraction $\alpha$ in Galactic globular clusters is relatively small, and usually does not exceed $\sim 10\%$ \citep{Milone2012}, with
only rare exceptions. For instance, \citet{Li17} found a much larger binary fraction for just three globular clusters ($\alpha = 0.6-0.8$). Open clusters (OCl), on the other hand,
host more significant fraction of binaries with $\alpha \geqslant 30\%$ \citep{Boni,KB,Sarro,alPer,Li17}. This percentage is, however, smaller than the one among field stars in the solar vicinity \citep{Duq1}. It has also been noted that the binary percentage increases at increasing a primary mass. This fact is often linked to the dynamical evolution of clusters \citep{KOP,Dorval}. 
Nevertheless, it does not seem to be universal since, for instance,
\citet{Patience} found  an increase of the companion-star fraction toward smaller masses in $\alpha$ Persei and Praesepe.

A fundamental quantity is the mass ratio $q$ distribution. Unfortunately, a consensus is still lacking.
According to \citet{Duq2}, the distribution of masses of the secondary in the field does not show a maximum
close to the unity. Instead, this distribution is continuously increased toward the low mass end. \citet{Fisher}, however,  found a
$q$ distribution peaking near $q=1$ for field stars. The same peak was found by \citet{Maxted} for the low-mass spectroscopic binaries in the young clusters around $\sigma ~Ori$ and $\lambda ~Ori$.  \citet{Ragh} support this point of view, showing that the mass ratio distribution shows a
preference for like-mass pairs, which occur more frequently in relatively close pairs.
\citet{RM} argue for a universal form of the $q$ distribution both for solar-type  and for M-dwarfs in the general Galactic field:

\begin{equation}\label{eq:power}
dN/dq \sim q^{\beta}
\end{equation}

\noindent
with the $\beta=0.25\pm0.29$ (flat within the errors ). Also, \citet{Milone2012} claim that in the interval $q\in[0.5,1.0]$ the distribution of $q$ is nearly flat, with few possible deviations
among Galactic globular clusters. \citet{Kouw} introduces two different $q$ distributions: a power-law (\ref{eq:power}) for $q\in[q_0,1]$ and
different $\beta$ values, and a Gaussian one

\begin{equation}\label{eq:gauss}
dN/dq \sim exp[-(q-\mu_q)^2/2\sigma_q^2]
\end{equation}

\noindent
for $q\in(0,1]$ with $\mu_q=0.23$ and $\sigma_q^2=0.42$.
According to \citet{Patience}, the $q$ distribution depends on the stellar mass interval: the higher mass systems reveal a decreasing
mass ratio distribution and the lower mass systems reveal a deficit of low mass ratio companions (see Fig.8 in \citet{Patience}). As a result,
the combined sample show the deficiency of $q>0.85$. However, a flat distribution is not ruled out (see Fig.6 in \citet{Patience}).

The $q$ distribution keeps a memory of the primordial binaries' properties. Some numerical experiments were carried out along this line \citep{Kroupa2011,Geller2013,PR}.
\citet{Geller2013} performed N-body simulations of the old open cluster
NGC 188 and showed that the distribution of orbital parameters for short-period  ($P<1000^d$) solar-type binaries would not be
changed significantly for several Gyrs of evolution. This fact means that observations of the present-day binaries even in the oldest open clusters
can bring essential information on the primordial binary population. On the other hand, \citet{PR} showed that while the overall binary fraction decreases,
the shape of the $q$ distribution remains unaltered during the evolution.
The presence of unresolved binaries in star clusters affects any estimate of their mass, both photometric
(via star counts) and dynamical (via velocity dispersion and the virial theorem). In the latter case,
if the sample of stars selected for velocity dispersion calculation
(through radial velocities) contains spectroscopic binaries, one can indeed artificially inflate the velocity dispersion, and hence increase the mass.
This point has been recently underlined by \citet{KdeG,Bianchini},  and by \citet{4337}.

When the cluster mass is evaluated through the luminosity function (LF) obtained via star counts, the mass estimate derived neglecting
unresolved binaries would result smaller than the actual mass. This is straightforward to show, since the mass of a binary system
is larger than the mass of a single star at the same magnitude due to the strong mass dependence of stars' luminosity
(approximately $(L/L_{\odot})\sim(M/M_{\odot})^4$ for the main-sequence stars, see Fig.7 on page 209 in \citet{CO}).

If a single star and  a binary system have the same
magnitude, their luminosities are also equal $L_s=L_1+L_2$, where suffix $s$ marks the single star, while $1$ and $2$ primary and secondary.
Therefore, $M_s^4=M_1^4+M_2^4$.  Instead, $(M_1+M_2)^4=M_1^4+M_2^4+4M_1M_2^3+6M_1^2M_2^2+4M_1^3M_2=
M_s^4+4M_1M_2^3+6M_1^2M_2^2+4M_1^3M_2$. Since all terms are positive, $(M_1+M_2)^4>M_s^4$, and $M_1+M_2>M_s$.

For example, the presence of unresolved binaries was taken into account by \citet{KB} to estimate the Praesepe cluster mass.
\citet{KB} found a binary fraction of $35\pm5$ percent in Preasepe and used a correction (multiplicative) factor of 1.35 for the cluster mass estimate. Following
them, the same correction was applied by \citet{profiles} to estimate of NGC 1502 stellar mass.
Unfortunately \citet{KB} provided little information on how they obtained the multiplicative correction factor 1.35, which leaves
room for further investigation.

To amend this, in this work we  present a novel approach and estimate the mass
of five open clusters of different age and metallicity, starting from  their LF.  In this case,
one can provide an independent estimate of this correction factor, and assess its
dependence both on binary fraction $\alpha$ and on $q$ distribution.

The layout of the paper is as follows.
Section 2 is devoted to the description of our approach and the associated algorithms; Section 3 contains our results for
NGC 1912, NGC 2099, NGC 6834, NGC 7142, and IC 2714. Section 4 is dedicated to a summary of our results,  provides the paper conclusions,
and discuss some future perspectives.

\section{Model and Algorithm}

Two ingredients are needed to derive the correction factor for applying to the photometric mass because of the presence of unresolved binaries.
The first one is the binary fraction. In this study, we adopt a binary fraction independent on magnitude.  A larger binary fraction for brighter
stars would not increase cluster mass significantly since bright, massive stars are typically only a few. We consider binary fraction in the range of 10-90\%.
The second one is the mass ratio $q$ distribution. We explore four different distribution functions for $q$:
\begin{itemize}
\item a $\delta$ function with $q=1$
\item a flat distribution function
\item a Gaussian distribution (\ref{eq:gauss}) as in \citet{Kouw}
\item a Gaussian distribution (\ref{eq:gauss}) with mode shifted to $q=1$ to reproduce \citet{Fisher}, \citet{Maxted}, and \citet{Ragh} functions.
\end{itemize}

\noindent
The last distribution was taken with $\mu_q=0.60$ and $\sigma_q^2=0.42$, the latter is the same as in \citet{Kouw}.

\citet{Kouw} summarised different methods of assignment of the mass values to the binary components, called by them as 'pairing' methods.
Our task is different because at the beginning we have the stellar magnitude of the binary and require the mass of each component.
The procedure described below could then be called 'luminosity limited pairing', following the terminology of \citet{Kouw}.

We use a quadratic mass-luminosity relation following \citet{Eker}:

\begin{equation}\label{eq:ML}
\log{L} = -(0.705\pm 0.041)(\log{M})^2 + (4.655\pm 0.042)(\log {M}) - (0.025\pm 0.010)\\
\end{equation}

\noindent
where $L$ is the luminosity, and $M$ the stellar mass (in solar units). Relation (3) refers to single main-sequence stars;
consequently, we assume that all unresolved binaries have their components at the main sequence and have not experienced mass
transfer. It is reasonable for stars below the MS turn-off . Then, only NGC 7142 with an age logarithm of 9.2 (see Table 1 below)
could contain a detectable number of binary stars after this stage of evolution. Nevertheless, even for stars above the turn-off in NGC 7142, we
could find only a few binary stars after the mass transfer. There could probably be five blue stragglers (see Figures 3 and 4 in \citet{7142})
and 1-2 evolved (yellow) stragglers among the upper part of CMD. \\
%If we use relation (\ref{eq:ML}) for these few stars, we introduce only a negligible
%inaccuracy into an estimate of the mass of binaries of the cluster. \\

We use the cluster luminosity function $\varphi(J)$ to count the number of stars in different magnitude intervals.
The luminosity functions are evaluated statistically with the use of 2MASS database \citep{2MASS},
that is we obtain the luminosity functions for the cluster region (``cluster plus field'')
and an equal area nearby reference field (``field'') and get the cluster luminosity function as a difference between "cluster plus field"
and ``field''. With this approach, we do not take into account a possible difference in the mass function between the cluster center and outskirts.
This procedure has been described in detail in \citet{LF,Pal1,4815,kernel,4337}.
The magnitude distribution is binned in
intervals $\Delta J$, and in each of them, we count number of stars, and then derive the number of binaries, using a binary fraction $\alpha$:

\begin{equation}
N= \int\limits_J^{J+\Delta J}{\varphi(J)dJ} \mbox{,} \mkern 25mu N_{b}=\alpha \int\limits_J^{J+\Delta J}{\varphi(J)dJ}
\end{equation}

\noindent
We round star numbers to integers and tune the number of intervals to get each bin occupied by at least one star. Fig.~1 illustrates the
process and  shows the luminosity function of NGC 7142 obtained as in \cite{kernel}.
For each magnitude bin, the mean magnitude is considered for further calculations.

\begin{figure}
   \centering
   \includegraphics[width=8truecm]{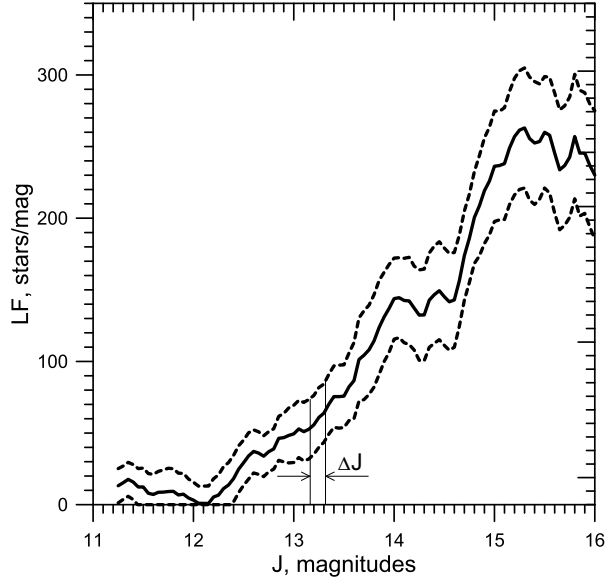}
   \caption{Luminosity function of NGC 7142 (solid line) and its $2\sigma$ confidence interval (dotted lines).
   A  J band bin is showed for illustration purposes. }
   \label{lf7142}
   \end{figure}

Stellar magnitudes are converted into luminosities with the use of the isochrone tables \citep{Padova} as follows.  An isochrone
corresponding to the cluster age is firstly selected. We took the cluster ages from \citet{LP}, but then refined them by comparing
with isochrones \citep{Padova} the cluster CMD for the probable cluster members selected from Gaia DR2 (\citet{2016gaia,2018gaia}) filtering by parallaxes and proper motions. Then, absolute magnitude is obtained from cluster distance modulus and colour excess. The set of adopted cluster data
is  listed in Table 1.
The photometric distances are then compared with Gaia DR2 distances, derived from parallaxes. We find that for distances closer than about 1.5 kpc,
photometric and Gaia distances agree exceptionally well. Beyond this distance, the figures provided by GAIA tend to be significantly larger
than the photometric ones. We tentatively impute such differences to the actual GAIA release, which is probably not very precise for large
distances. Future releases will surely alleviate these differences.

Then, the star mass and the luminosity value are extracted from the isochrone table corresponding to each cluster age.
Table 1 contains the limits in the star masses covered by our luminosity functions:
column 7 contains the minimum mass (it corresponds to J=16 mag with exception to NGC 6834, where the minimum mass corresponds to J=15.9 mag;
these magnitudes, in turn, correspond to the completeness limit of the 2MASS data) and column 8 contains the maximum mass.
Stars with masses close to the upper mass limit have been evolved from the main sequence. Due to this reason we use another isochrone table
with an age of $4\cdot10^7$ years to determine the luminosity of the evolved stars at the main sequence stage with the same mass as evolved star
mass.
For each binary, the following system of equations holds:

\begin{equation}
\left\{
\begin{array}{lcl}
L & = & L_1+L_2\\
\log{L_1} & = & -0.705(\log{M_1})^2  + 4.655(\log {M_1}) - 0.025\\
\log{L_2} & = & -0.705(\log{M_2})^2  + 4.655(\log {M_2}) - 0.025\\
q & = & M_2 / M_1\\
\end{array}
\right.\label{eq:system}
\end{equation}

\noindent
where L is luminosity of binary star, $L_1$ and $L_2 $ are luminosities of the binary components, $M_1$ and $M_2$ are masses of the
primary and secondary components of the binary star, respectively. For each binary star, we extract mass ratio $q$ from the component mass ratio
distribution from Monte-Carlo simulations.

\begin{deluxetable}{rccccccc}
\tablecaption{Star cluster characteristics\label{tab:charact}. }
\tablehead{
\colhead{Cluster} & \colhead{$\log t$}  & \colhead{$(m-M)_0^1$} & \colhead{$E(B-V)^1$} & \colhead{$d_{PHOT}$} & \colhead{$d_{GAIA}$} & \colhead{$M_{min}$} &  \colhead{$M_{max}$}\\
\colhead{} & \colhead{$t$ in years} & \colhead{mag}  & \colhead{mag}  &\colhead{pc}  & \colhead{pc} & \colhead{$M_{\odot}$}  & \colhead{$M_{\odot}$} \\
}
\colnumbers
\startdata
  IC 2714 & 8.6 & 10.48 & 0.34 & 1250 & 1390 & 0.73 & 2.82 \\
 NGC 1912 & 8.3 & 10.29 & 0.25 & 1140 & 1150 & 0.68 & 3.60 \\
 NGC 2099 & 8.7 & 10.74 & 0.30 & 1410 & 1510 & 0.76 & 2.77 \\
 NGC 6834 & 7.9 & 11.59 & 0.71 & 2080 & 3570 & 1.07 & 5.12 \\
 NGC 7142 & 9.2 & 11.25 & 0.39 & 1780 & 2600 & 0.87 & 1.80 \\
\enddata
\tablecomments{1 - \citet{LP}.}
\end{deluxetable}

Let be $x=\log{M_1}$, $a=-0.705$ , $b=4.655$, and $c=- 0.025$. After some algebra, luminosity reads:

\begin{equation}
\ln{L} = \ln{10}\cdot(ax^2+bx+c)+\ln{(1 + e^{\ln{10}\cdot(a(\log {q})^2+\log {q}(b+2ax))})}
\end{equation}

\noindent
The goal is to define $x$, so that we build  up a function $f(x)$, which is equal to zero when a solution to the system (\ref{eq:system}) is found:

\begin{equation}
f(x)=\ln{10}\cdot (ax^2+bx+c)+\ln{(1 + e^{\ln{10}\cdot(a(\log {q})^2+\log {q}(b+2ax))})} - \ln{L}
\end{equation}

To solve this equation we use the Newton-Raphson method as

\begin{equation}
x_{k+1}=x_k-f(x_k)/f'(x_k) \qquad \mbox{, where} \qquad f'(x)=df(x)/dx
\end{equation}

\noindent
until the difference $|x_{k+1} - x_k|$ reaches the requested accuracy.

The Newton-Raphson method converges only if certain conditions are met.
Firstly, one needs to choose initial trial values, which are not too far from the root. Therefore we
build a {\it for-loop} with intervals of the mass $[0.08 ;10] M_{\odot}$ , (or $x \in [-1.097,1]$) with a small increase. Loop ends when we find those $x_i$, which give $f(x_i)\cdot f(x_{i+1}) <0$; this implies that the root is in the interval $x \in [x_i,x_{i+1}]$. We then
consider $x_i$ as a starting point for iteration.
Secondly, function $f(x)$ should be smooth in its domain; this is easy to prove, being $f(x)$ a combination of smooth functions.

The final $x_k$ will be the solution of the equation and, in turn, $M_1=10^x$ the value for the primary component mass from the system (\ref{eq:system}).
Hence, we can define the secondary component mass $M_2$ from the fourth equation of the system (\ref{eq:system}), and, finally,  the total mass of the binary star $M_1+M_2$.
The described procedure is repeated for all $N_b$ stars to eventually derive the total mass of binaries in the interval $J\in [J;J+\Delta J]$.
When extended to all magnitude bins, the procedure yields the total mass of the cluster binaries $M_b$ in these bins.

Finally, to define the mass of the cluster, we need to find the mass stored in single stars $M_s$ (whose number is $N_s=N-N_b$ in each
magnitude interval). For these stars, we use an isochrone table, where we determine the mass according to the magnitude and the cluster
parameters from Table 1 (see a description of the procedure above). As a result, we obtain the cluster mass $M = M_b+M_s$
in the considered magnitude interval.

Let us now define $M_{wob}$ as the cluster mass obtained assuming that all stars are single. Then
the ratio $M/M_{wob}$ would naturally give the cluster mass increment due to unresolved binaries.

%\newpage

\begin{figure}
\gridline{\fig{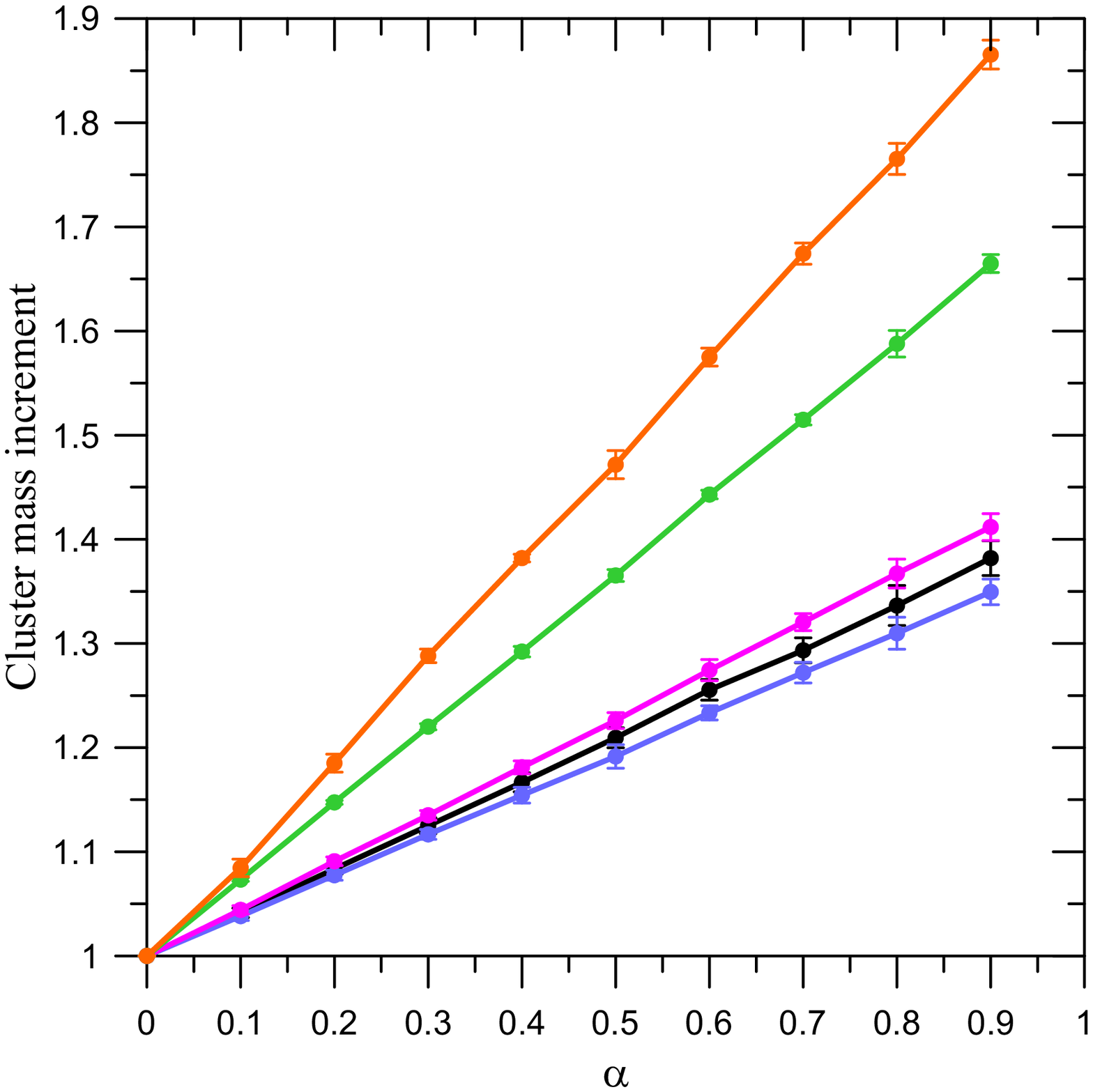}{0.3\textwidth}{(a)}
          \fig{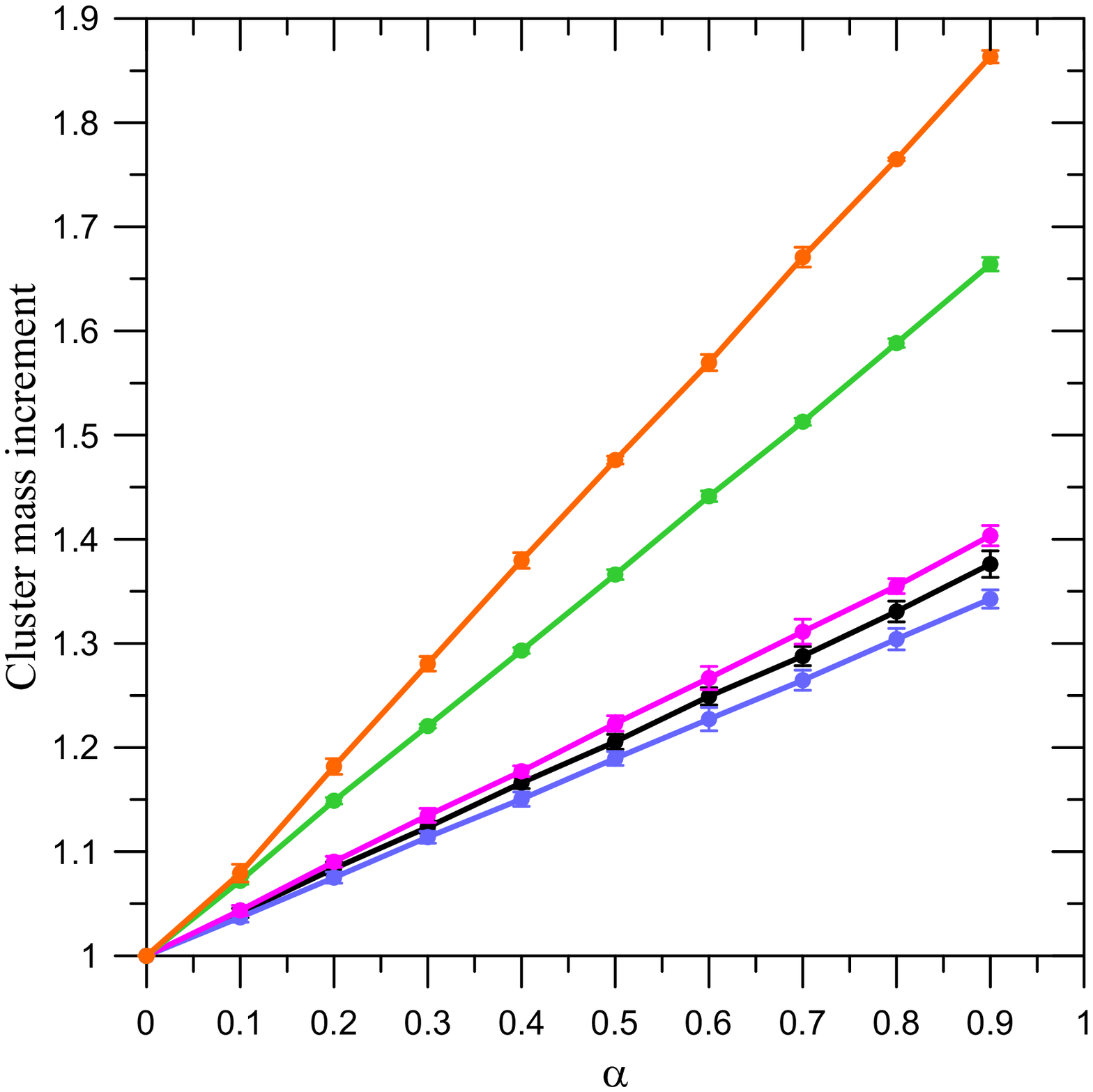}{0.3\textwidth}{(b)}
          \fig{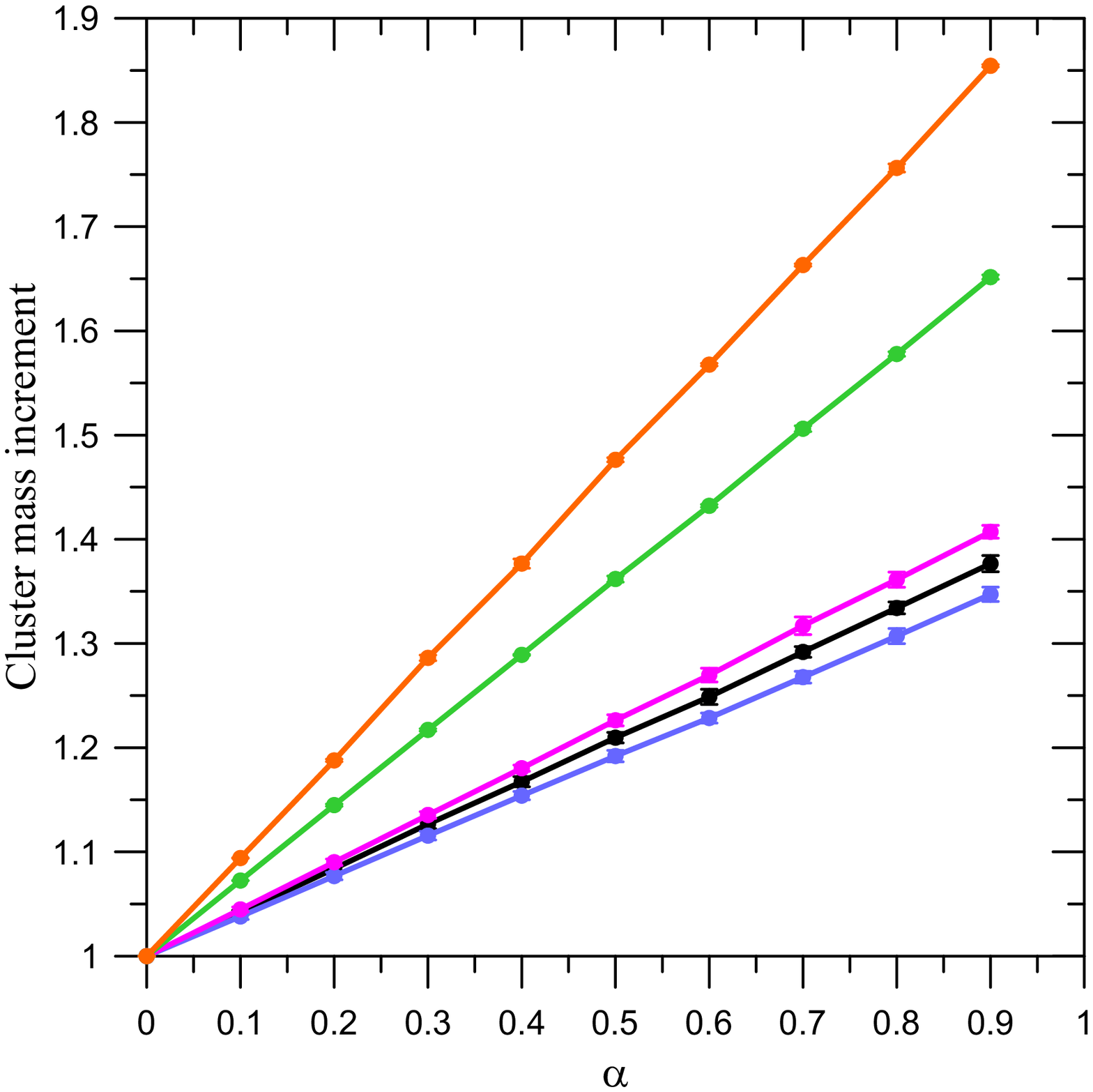}{0.3\textwidth}{(c)}}
\gridline{\fig{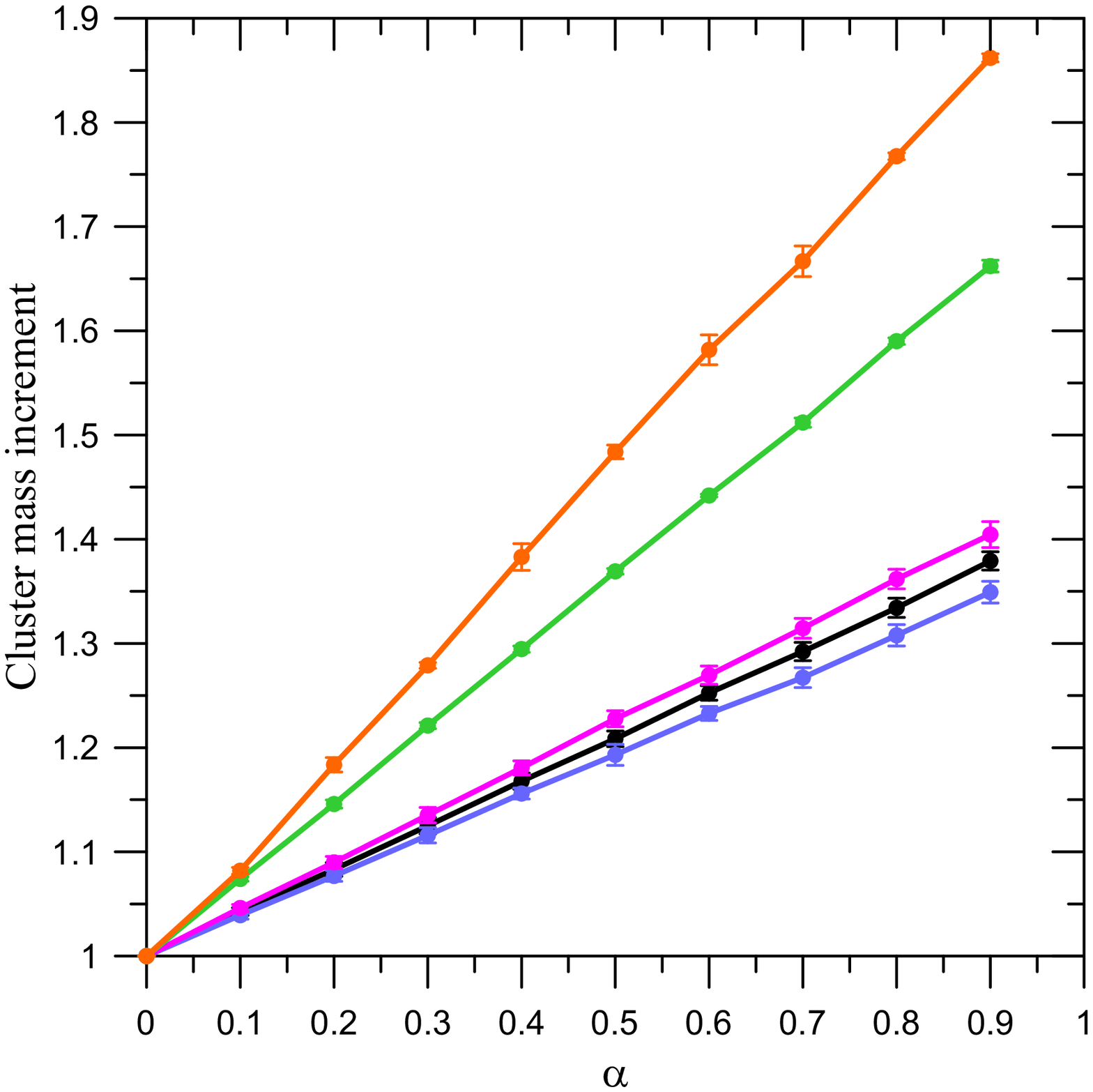}{0.3\textwidth}{(d)}
          \fig{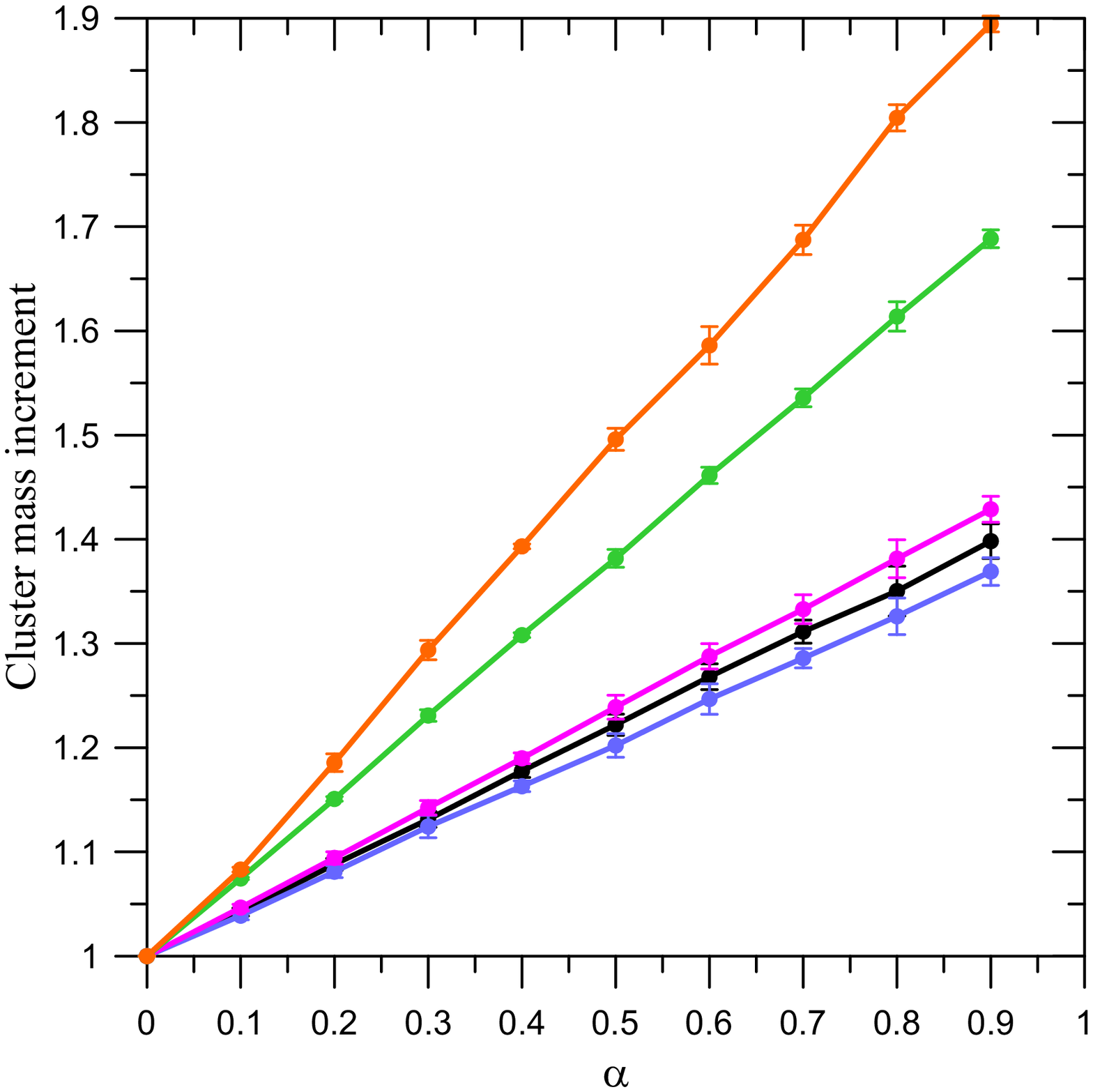}{0.3\textwidth}{(e)}}
\caption{The dependence of the cluster mass increment on the binary fraction $\alpha$ with different assumptions on the
binary component mass ratio $q$ distribution. Green: equal components, orange: equal components with taking into account
the multiple (triple and quadruple) systems, black: flat
distribution, electric blue: gauss distribution with a maximum close to zero, magenta: gauss
distribution with a maximum close to unity. (a) IC 2714; (b) NGC 1912; (c) NGC 2099; (d) NGC 6834; (e) NGC 7142.}
\label{increments}
\end{figure}

%\startlongtable
\begin{deluxetable}{lccccc}
\tablecaption{Linear approximation $y=A+B\alpha$ for the cluster mass increment dependence on the binary fraction \label{tab:approx}}
\tablehead{
\colhead{$q$ distribution} & \colhead{Cluster} & \colhead{A}  & \colhead{B} & \colhead{$\chi^2$} & \colhead{Q} \\
\colhead{model} & \colhead{} & \colhead{}  & \colhead{}  & \colhead{}   & \colhead{}  \\
}
\colnumbers
\startdata
Equal component    &  IC 2714 & 1.000 $\pm$ 0.002 & 0.736 $\pm$ 0.005 & 0.673  &  1.000 \\
masses             & NGC 1912 & 1.000 $\pm$ 0.002 & 0.735 $\pm$ 0.005 & 1.162  &  0.997 \\
                   & NGC 2099 & 1.000 $\pm$ 0.000 & 0.722 $\pm$ 0.001 & 1.639  &  0.990 \\
                   & NGC 6834 & 1.000 $\pm$ 0.002 & 0.736 $\pm$ 0.004 & 1.025  &  0.998 \\
                   & NGC 7142 & 0.997 $\pm$ 0.001 & 0.773 $\pm$ 0.006 & 1.263  &  0.996 \\
\hline
Equal component    &  IC 2714 & 0.994 $\pm$ 0.005 & 0.968 $\pm$ 0.012 & 1.690  &  0.989 \\
masses with triple & NGC 1912 & 0.991 $\pm$ 0.004 & 0.968 $\pm$ 0.006 & 2.316  &  0.970 \\
and quadruple      & NGC 2099 & 0.999 $\pm$ 0.000 & 0.949 $\pm$ 0.001 & 5.274  &  0.728 \\
systems            & NGC 6834 & 0.987 $\pm$ 0.003 & 0.975 $\pm$ 0.005 & 5.582  &  0.694 \\
                   & NGC 7142 & 0.983 $\pm$ 0.003 & 1.020 $\pm$ 0.007 & 6.496  &  0.592 \\
\hline
Flat               &  IC 2714 & 0.999 $\pm$ 0.004 & 0.423 $\pm$ 0.011 & 0.181  &  1.000 \\
distribution       & NGC 1912 & 1.000 $\pm$ 0.004 & 0.414 $\pm$ 0.009 & 0.227  &  1.000 \\
                   & NGC 2099 & 1.000 $\pm$ 0.002 & 0.417 $\pm$ 0.005 & 0.194  &  1.000 \\
                   & NGC 6834 & 1.000 $\pm$ 0.003 & 0.419 $\pm$ 0.008 & 0.185  &  1.000 \\
                   & NGC 7142 & 0.998 $\pm$ 0.004 & 0.447 $\pm$ 0.011 & 0.190  &  1.000 \\
\hline
Gaussian           &  IC 2714 & 1.000 $\pm$ 0.004 & 0.389 $\pm$ 0.010 & 0.086  &  1.000 \\
distribution       & NGC 1912 & 0.999 $\pm$ 0.004 & 0.381 $\pm$ 0.009 & 0.061  &  1.000 \\
$\mu_q=0.23$       & NGC 2099 & 1.000 $\pm$ 0.003 & 0.384 $\pm$ 0.006 & 0.273  &  1.000 \\
                   & NGC 6834 & 1.000 $\pm$ 0.003 & 0.386 $\pm$ 0.008 & 0.256  &  1.000 \\
                   & NGC 7142 & 0.998 $\pm$ 0.004 & 0.411 $\pm$ 0.010 & 0.167  &  1.000 \\
\hline
Gaussian           &  IC 2714 & 0.998 $\pm$ 0.003 & 0.459 $\pm$ 0.009 & 0.218  &  1.000 \\
distribution       & NGC 1912 & 1.000 $\pm$ 0.004 & 0.446 $\pm$ 0.009 & 0.265  &  1.000 \\
$\mu_q=0.60$       & NGC 2099 & 1.000 $\pm$ 0.002 & 0.452 $\pm$ 0.006 & 0.073  &  1.000 \\
                   & NGC 6834 & 1.001 $\pm$ 0.003 & 0.450 $\pm$ 0.009 & 0.161  &  1.000 \\
                   & NGC 7142 & 0.999 $\pm$ 0.003 & 0.478 $\pm$ 0.010 & 0.058  &  1.000 \\
\enddata
\end{deluxetable}

\section{Results for the program clusters}

In this work, we start from the luminosity function of five open clusters: IC 2714, NGC 1912, NGC 2099, NGC 6834 and NGC 7142 obtained by star-counts  with 2MASS
as described above.

For each cluster,  we repeated the procedure described in the previous Section up to 30 times both for cluster luminosity function and for
boundaries of the LF confidence interval. This procedure allowed us to evaluate the scatter of the mass increment factors.
We explored the whole parameter space made of binary fraction $\alpha$ and mass ratio $q$ distribution to quantify the spread in the estimates of the cluster mass when unresolved binaries are taken into account.

We considered two cases of equal mass components. The first case is when we take into account binary systems only. In the second case, we
also take into account the multiple (triple and quadruple) systems following  \citet{multi}, who found for systems with multiplicity of
1:2:3:4:5 (1 means single star) the relative abundance ratio of 54:33:8:4:1. It is worthwhile because at distances
of $\sim$ 1 kpc a hierarchical triple of separation $\sim$ 100 au has an angular separation of about 0.1 arcsec, then
a triple system or a ``binary of binaries'' could be missed, just like tight
unresolved binaries.

Fig.2 shows the dependence of the cluster mass increment on the binary fraction for the five clusters. Each
panel corresponds to a cluster, and different colors are used to indicate the various  $q$ distributions. At first glance, one can easily see that
the equal mass component model significantly deviates from the other models, which do not appear much different.

\citet{KB} found the cluster mass increment value of 1.35 for a binary fraction of 0.35.
According to our study, the increment value should be between 1.10 and 1.15 for {\it realistic}
$q$ distribution (see Fig.2). However, taking into account the possible presence of the multiple (triple and quadruple) systems in the
cluster would increase the value of the increment on the average 1.32 times for the case of equal components.  Then the value of 1.35 for
the cluster mass increment found by \citet{KB} for the Praesepe cluster is reasonable.
We fitted the dependencies of the increment on the binary fraction via linear regression, and provide fitting formulae in Table 2.
The columns of Table 2 are: the binary components mass ratio model, the cluster, the coefficients
$A$ and $B$ of the linear regression $y=A+B\alpha$ (where $y$ is the cluster mass increment, and $\alpha$ is the binary fraction), the $\chi^2$
of the fit, and the goodness-of-fit probability Q (\cite{NumRec}). Coefficient $A$ does not differ significantly from the unity in virtually all cases. The coefficients
$B$ for the clusters lie within the limits of the $q$ distribution model (except for NGC 7142, the oldest one).
This fact demonstrates that the shape of the luminosity function
does not affect the dependence of the cluster mass increment on the binary fraction $\alpha$ significantly.

 The luminosity functions used in the present work are limited in magnitude because of the completeness limit of 2MASS. Therefore we miss stars with masses lower than the limit listed in the 7th column of Table 1. How can the missing low-mass stars affect our results? We consider the
binary fraction $\alpha$ independent of the stellar magnitude. In such a case, the cluster mass increment should be independent of the
magnitude (and the mass) limit. In order to make this suggestion more solid  we performed the following experiment. For NGC 2099 we calculated the mass increment for
a set of limiting magnitudes $J=14, 15, 16$ mag in the case of flat $q$ distribution. It turned out that the mass increment slightly increases with
the limiting magnitude. For instance, for $\alpha=0.8$ $y=1.322\pm0.013$ for $J_{lim}=14$ mag, $y=1.328\pm0.009$ for $J_{lim}=15$ mag, and
$y=1.334\pm0.006$ for $J_{lim}=16$ mag. If the binary fraction increases with the stellar magnitude, the cluster mass increment would  most probably increase
with the stellar magnitude. If the binary fraction decreases with the stellar magnitude, we would expect the cluster mass
increment being independent on the stellar magnitude or even decreasing with the stellar magnitude.

In any case, we underline that even applying the mass increment one would not get the total mass of the cluster but only slightly improve a lower
limit estimate of it.

\section{Conclusions}
In this work, we attempt to quantify the increase of the cluster mass estimate --- obtained by star counts --- produced by the presence of
unresolved binaries. The results are illustrated in Fig.2 and summarised in Table 2.

The most relevant results of this study are:

\begin{itemize}

\item the dependence of the cluster mass increment on binary fraction is linear in most cases.

\item the dependence of the  cluster mass increment on the binary fraction $\alpha$ does not vary significantly for the {\it realistic} $q$ distributions
considered here. We checked three {\it  realistic} distributions: a Gaussian distribution (\ref{eq:gauss}) with
$\mu_q=0.23$, a flat distribution, and  a Gaussian distribution (\ref{eq:gauss}) with $\mu_q=0.60$. An inspection of Fig. 2 and Table 2
shows that the closer the distribution mode to unity, the higher is the expected cluster mass increment .

\item  the dependence of the cluster mass increment on
the binary fraction $\alpha$ within the limits of a specific  $q$ distribution model does not differ substantially among the selected
clusters (except for NGC 7142, the oldest one). Then
we can safely conclude that the form of the luminosity function does not affect this dependence considerably.

\item for the particular case of a binary fraction $\alpha=0.35$ the cluster mass increment is confined between 1.10 and 1.15
(for {\it realistic}
$q$ distributions, see Fig.2). However, taking into account the possible presence of the multiple (triple and quadruple) systems in the
cluster would increase the value of the increment (in the mean 1.32 times for the case of equal components).  Then the value of 1.35 for
the cluster mass increment for the Praesepe cluster obtained by \citet{KB} is reasonable.

\end{itemize}

\noindent
Our results will help to improve the estimate of the mass of clusters containing unresolved binary stars in the broad range of the binary
ratios $\alpha$ and with different assumptions on the distribution of the binary component mass ratio $q$.

\acknowledgments
The work of Anton F. Seleznev and Vladimir M. Danilov was partly supported by the Ministry of Education and Science
(the basic part of the State assignment, RK no. AAAA-A17-117030310283-7). The work of Anton F. Seleznev and Vladimir M. Danilov
was supported also by the Act no. 211 of the Government of the Russian Federation, agreement no. 02.A03.21.0006.

This work has made use of data from the European Space Agency (ESA) mission
{\it Gaia} (\url{https://www.cosmos.esa.int/gaia}), processed by the {\it Gaia}
Data Processing and Analysis Consortium (DPAC,
\url{https://www.cosmos.esa.int/web/gaia/dpac/consortium}). Funding for the DPAC
has been provided by national institutions, in particular the institutions
participating in the {\it Gaia} Multilateral Agreement.

This publication makes use of data products from the Two Micron All Sky Survey, which is a joint project of the University of Massachusetts
and the Infrared Processing and Analysis Center/California Institute of Technology, funded by the National Aeronautics and Space Administration
and the National Science Foundation.

{}

\end{document}